\documentclass[pre,twocolumn,showpacs,superscriptaddress,floatfix]{revtex4}
\usepackage{graphicx}
\usepackage{epsfig}
\usepackage{bm}

\begin{document}
\title{Intensity Statistics of Random Signals in Gaussian Noise}

\author{A.A.~Chabanov}
\affiliation{Department of Physics and Astronomy,
University of Texas at San Antonio, San Antonio, TX 78249, USA}

\date{24 June, 2009}

\begin{abstract}

The intensity statistics of signals in the presence of Gaussian noise is obtained by studying the model of a random signal plus a random phasor sum. The additive Gaussian noise is shown to result in a Bessel transform of the probability density of the signal intensity. The transformation of the intensity statistics can generally be applied to a mixture of independent random signals, one of which being a complex-valued Gaussian random process. It is used to retrieve intensity statistics of microwave pulsed transmission from Gaussian noise at long time delays.

\end{abstract}


\maketitle
\section{Introduction}

The signal available at the output of a radio measurement system is never an entirely accurate indicator of the quantity to be measured. The accuracy of the measurement depends on the amount of noise compared to the level of signal, or signal-to-noise ratio. Signal noise can result from a variety of causes, both man-made and natural. In most cases, however, it is natural additive noise which is  the limiting factor for signal detection. This may include noise radiation from the sky picked up at the antenna, Johnson-Nyquist noise and shot noise generated in the circuitry of the receiver, etc. Usually the natural additive noise can be represented mathematically as a Gaussian random process, hence the term Gaussian noise. The problems of detecting signals in Gaussian noise and of estimating parameters of signals in Gaussian noise have been studied by means of detection theory. There exists an extensive literature on this subject (see, for example, \cite{SDT} for review).

Our interest in the statistics of random signals in Gaussian noise arose from measurements of electromagnetic waves transmitted through random media. Wave transport in the presence of disorder can be characterized by the degree of nonlocal intensity correlation, which reflects the closeness to the Anderson localization transition (see, for example, \cite{ShengBook} and \cite{RMP}). The presence of long-range correlation of intensity within a sample leads to enhanced fluctuations of total transmission over the value predicted if correlation of intensity were short range, as is correlation of the field. The occurrence of enhanced transmission fluctuations can be seen in an ensemble of quasi-1D random samples, in which the sample length is much greater than the diameter of its cross section \cite{Marin}. In pulsed transmission measurements, the variance of transmission fluctuations normalized to the ensemble-averaged transmission increases with time delay from an exciting pulse \cite{StDy} while the decay rate of the average intensity within the sample decreases \cite{AvDy}, reflecting two related effects: the increasing impact of localization and the growing weight of long-lived electromagnetic quasi-modes. At long times, however, the decaying transmitted intensity becomes comparable to Gaussian noise then affecting the intensity statistics so that the measured variance of total transmission is no longer the localization parameter. To study the dynamics of transport at long time delays, the impact of Gaussian noise must therefore be determined.

To solve this problem, we consider the model of a random signal plus noise in which noise is represented by a random phasor sum with circular Gaussian statistics. We find that the addition of complex-valued Gaussian noise results in a Bessel transform of the probability density of the signal intensity. Depending on the physical problem under consideration, the solution can be used to find the intensity statistics of signal in the presence of Gaussian noise, to retrieve the intensity statistics of signal from Gaussian noise given the noise intensity, or to determine the noise intensity given the intensity distribution of the signal. We shall use it to determine the probability density of pulsed transmitted intensity in Gaussian noise at long time delays. More generally, the solution can be applied to a mixture of independent random signals, one of which being a complex-valued Gaussian random process. An important example is the stationary field of a disordered cavity coupled to the environment, which can be represented by superposition of a standing wave (an eigenstate) and a traveling wave associated with the energy leaking out of the system \cite{Pnini}, the former playing the role of `signal' and the latter playing the role of `Gaussian noise'.

\section{Statistical model of a random signal plus Gaussian noise}

We consider the model of a random signal plus Gaussian noise, $\bm{E}=\bm{E}_{\text s}+\bm{E}_{\text n}$, in which Gaussian noise, $\bm{E}_{\text n}$, is represented by a random phasor sum with circular Gaussian statistics \cite{Goodman}. The probability density of the real and imaginary parts of $\bm{E}_{\text n}$, $r_{\text n}$ and $i_{\text n}$, respectively, is a Gaussian with width $\sigma$,
\begin{equation}
P(\bm{E}_{\text n})\equiv P(r_{\text n},i_{\text n})={1\over 2\pi\sigma^2}\exp\left(-{r^2_{\text n}+i^2_{\text n}\over 2\sigma^2}\right),
\end{equation}
and the moments,
\begin{equation}
\langle r^n_{\text n}\rangle=\langle i^n_{\text n}\rangle=\left\{
\begin{array}{cl}
1\cdot 3\cdot 5\cdots(n-1)\,\sigma^n\,, & \text{$n$ even} \\
0\,, & \text{$n$ odd}
\end{array}
\right.
\end{equation}
where $\langle ...\rangle$ represents the average over an ensemble of realizations. Because $\bm{E}_{\text s}$ and $\bm{E}_{\text n}$ are statistically independent, the probability density of the resultant field $\bm{E}$ is simply the convolution, $P(\bm{E})=P(\bm{E}_{\text s})*P(\bm{E}_{\text n})$. Finding statistics of the intensity of the resultant field, $I=|\bm{E}|^2$, however, is more involved.

The Gaussian statistics of $r_{\text n}$ and $i_{\text n}$ results into the exponential probability density of the noise intensity, $I_{\text n}=|\bm{E}_{\text n}|^2=r^2_{\text n}+i^2_{\text n}$,
\begin{equation}
P(I_{\text n})={1\over\langle I_{\text n}\rangle}\exp\left(-{I_{\text n}\over\langle I_{\text n}\rangle}\right),
\end{equation}
where $\langle I_{\text n}\rangle=2\sigma^2\equiv D$ is the average noise intensity that we shall denote by $D$. The moments of the noise intensity follow from (3) as $\langle I^n_{\text n}\rangle=n!\,D^n$. The signal intensity, $I_{\text s}$, can be written as $I_{\text s}=|\bm{E}_{\text s}|^2=r^2_{\text s}+i^2_{\text s}$, where $r_{\text s}$ and $i_{\text s}$ are the real and imaginary parts of $\bm{E}_{\text s}$, respectively. The moments of $I_{\text s}$ are given by
\begin{equation}
\langle I^n_{\text s}\rangle=\!\!\int_0^\infty\!\!\! dI_{\text s}\,I^n_{\text s}\,P(I_{\text s}),
\end{equation}
where $P(I_{\text s})$ is the probability density of the signal intensity.

To find the probability density of the intensity of the resultant field, we first calculate its moments $\langle I^n\rangle$. Expressing $I^n$ in terms of the real and imaginary parts of the signal and noise, taking the average and using (2), we arrive at
\begin{equation}
\langle I^n\rangle=\sum_{k=0}^n {(n!)^2\over (k!)^2(n-k)!}\,\langle I_{\text s}^{k} \rangle D^{n-k}.
\end{equation}
From (5), the average and the variance are $\langle I\rangle=\langle I_{\text s}\rangle+D$ and $\text{var}(I)=\text{var}(I_{\text s})+2\langle I_{\text s}\rangle D+D^2$, respectively.

From the moments $\langle I^n\rangle$ the characteristic function and probability density of $I$ can be obtained \cite{Goodman}. Here these are derived using the characteristic function $M_{I_{\text s}}$ of the signal intensity $I_{\text s}$,
\begin{equation}
M_{I_{\text s}}(p)\equiv\langle\exp(-pI_{\text s})\rangle=\sum_{n=0}^{\infty} {(-p)^n\over n!}\langle I^n_{\text s}\rangle.
\end{equation}
The moments $\langle I^n_{\text s}\rangle$ can be deduced from (5) and written as
\begin{equation}
\langle I^n_{\text s}\rangle=(-1)^nn!D^n\langle L_n(I/D)\rangle,
\end{equation}
where $L_n$ is the Laguerre polynomial of order $n$. Substituting (7) into (6) and by making use of a generating function of Laguerre polynomials \cite{Arfken}, we obtain
\begin{equation}
M_{I_{\text s}}(p)={1\over 1-pD}\left\langle\exp\!\left(-{pI\over1-pD}\right)\right\rangle,\,\,|pD|<1.
\end{equation}
By changing variables, $pD=sD/(1+sD)$, Eq.~(8) can be written as
\begin{equation}
{1\over1+sD}M_{I_{\text s}}\!\left({s\over 1+sD}\right)=\left\langle\exp(-sI)\right\rangle\equiv M_I(s),
\end{equation}
where $M_I(s)$ is the characteristic function of the intensity $I$. $P(I)$ is related to $M_I(s)$ in the usual way,
\begin{equation}
P(I)=\!\!\int_{-i\infty}^{i\infty}{ds\over 2\pi i}\exp(sI)M_I(s).
\end{equation}
Substituting $M_I(s)$ of (9) into (10), we obtain
\begin{equation}
P(I)=\!\!\int_{-i\infty}^{i\infty}\!{ds\over 2\pi i}\,e^{sI}\!
\!\int_0^{\infty}\!\!\!{dI_{\text s}\over 1+sD}\,P(I_{\text s})\,e^{-sI_{\text s}/(1+sD)}.
\end{equation}
Changing the order of integration and integrating over $s$, we arrive at
\begin{equation}
P(I)={1\over D}\,e^{-I/D}\!\!\int_0^{\infty}\!\!\!dI_{\text s}\,P(I_{\text s})\,e^{-I_{\text s}/D}\,
{\cal I}_0\!\left({2\sqrt{II_{\text s}}\over D}\right),
\end{equation}
where ${\cal I}_0$ is a modified Bessel function of the first kind of zero order. Eq.~(12) is the main result of the paper and represents the transformation of the probability density of the signal intensity in the presence of Gaussian noise with the average intensity $D$. Some examples are in order. For a constant signal with intensity $A$, for which $P(I_{\text s})=\delta(I_{\text s}-A)$, we obtain from (12)
\begin{equation}
P(I)={1\over D}\,\exp\!\left(-{I+A\over D}\right)\,{\cal I}_0\!\left({2\sqrt{IA}\over D}\right),
\end{equation}
which is in agreement with the result of the model of a constant phasor plus a random phasor sum \cite{Goodman,Kogan,Chabanov}. In the case when $\bm{E}_{\text s}$ is itself a random phasor sum, $P(I_{\text s})=\exp\!\left(-I_{\text s}/\langle I_{\text s}\rangle\right)/\langle I_{\text s}\rangle$. Then, as may be expected, Eq.~(12) yields $P(I)=\exp\!\left(-I/\langle I\rangle\right)/\langle I\rangle$, where $\langle I\rangle=\langle I_{\text s}\rangle+D$. Finally, in the case of a disordered cavity in the absence of non-proportional damping \cite{Weaver}, $P(I_{\text s})=\exp\!\left(-I_{\text s}/2\langle I_{\text s}\rangle\right)/\sqrt{2\pi I_{\text s}\langle I_{\text s}\rangle}$, which is the Porter-Thomas distribution \cite{Pnini}. We then obtain from (12) the probability density, $P(x=I/\langle I\rangle)$, in the crossover from closed to open system,
\begin{equation}
P(x)={1\over\sqrt{\delta}}\,\exp\!\left(-{x\over\delta}\right)\,{\cal I}_0\!\left({x\sqrt{1-\delta}\over\delta}\right),
\end{equation}
where $\delta=(2D\langle I_{\text s}\rangle+D^2)/(\langle I_{\text s}\rangle+D)^2$, in agreement with Ref.~\onlinecite{Pnini}.

The equation (12) can be inverted to yield the probability density of the signal intensity,
\begin{equation}
P(I_{\text s})={1\over D}\,e^{I_{\text s}/D}\!\!\int_0^{\infty}\!\!\!dI\,P(-I)\,e^{-I/D}\,
J_0\!\left({2\sqrt{II_{\text s}}\over D}\right).
\end{equation}
However, Eq.~(15) is not particularly useful to retrieve $P(I_{\text s})$ from Gaussian noise with the average intensity $D$, because it requires a continuation of $P(I)$ to negative $I$, which is not available from the measurement. Instead, $P(I_{\text s})$ can be found by solving the equation (12). In the next section, we determine the probability density of pulsed transmitted intensity in Gaussian noise at long time delays.

\section{Intensity statistics of pulsed transmission in Gaussian noise}

Here the results of statistical model of the previous section are used to retrieve the intensity statistics of pulsed microwave transmission through random media from Gaussian noise at long delay times \cite{StDy,AvDy}. Accurate measurements of the time-resolved statistics of pulsed transmission are absolutely essential for systematic study of wave transport in the presence of disorder. Spectral measurements of the field transmission coefficient of microwave radiation were made in an ensemble of random dielectric samples, as described in \cite{StDy,AvDy}. The response to a pulse with a Gaussian temporal envelope at carrier frequency $\nu_{0}$ is obtained by Fourier transforming the product of the field transmission spectrum and a Gaussian spectral function of width $\delta\nu$. The field of the temporal response is squared to give the intensity $I(t)$ for each sample configuration. The average intensity $\langle I(t)\rangle$ is found by averaging over the ensemble of realizations. The result is shown on a logarithmic scale as the black solid line in Fig.~1a. The noise in the transmitted field manifests itself as a constant background in Fig.~1a. The analysis of the probability density $P(r,i)$ in the negative time before the pulse, i.e., when $\bm{E}=\bm{E}_{\text n}$, shows that $P(r_{\text n},i_{\text n})$ is a circular Gaussian. The average pulsed transmitted intensity $\langle I_{\text s}(t)\rangle$ is then $\langle I_{\text s}(t)\rangle=\langle I(t)\rangle-D$, where $D$ is the constant background in Fig.~1a. Once this background is subtracted (blue dashed line), the dynamic range is significantly enhanced.
\begin{figure}[!]
\includegraphics[width=3.5in]{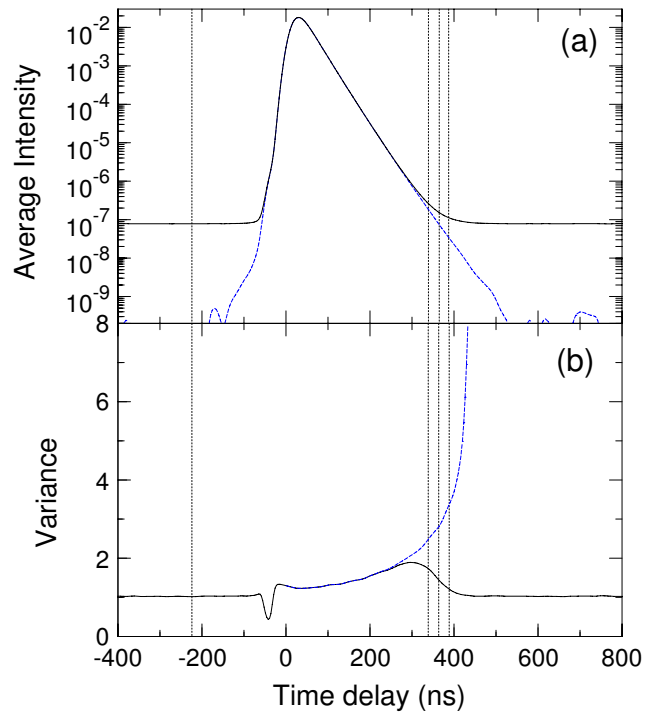}
\caption{(color on line). Measurements of the time-resolved statistics of pulsed microwave transmission in an ensemble of random dielectric samples. The average (a) and the variance normalized to the average (b) of the measured intensity $I(t)$ (black solid lines) and the transmitted intensity $I_{\text s}(t)$ (blue dashed lines) deduced from Eq.~(7).}
\end{figure}

The variance of normalized intensity, $\mathrm{var}[I(t)/\langle I(t)\rangle]$, is shown as the black solid line in Fig.~1b. In the pulsed measurement the variance of normalized transmitted intensity is expected to increase with time delay from an exciting pulse \cite{StDy} as increasingly more of energy within the medium is stored in long-lived localized modes. In Fig.~1b, in contrast, the variance of normalized intensity falls at long times to a value of unity, reflecting the increasing impact of Gaussian noise. The variance of transmitted intensity follows from (7) as $\text{var}[I_{\text s}(t)]=\text{var}[I(t)]-2\langle I(t)\rangle D+D^2$. The evolution with time of the normalized variance, $\mathrm{var}[I_{\text s}(t)/\langle I_{\text s}(t)\rangle]$, is shown as the blue dashed line in Fig.~1b.

As it follows from Fig.~1, the probability distribution of the transmitted intensity is increasingly affected by Gaussian noise at long times. The probability density $P(I/\langle I\rangle)$ for the time delays $t=338$, 364, and 389 ns, indicated by vertical dashed lines in Fig.~1, are shown as the black solid curves in Fig.~2. At these time delays, the relative amount of noise is $D/\langle I\rangle=0.30$, 0.51, and 0.72, respectively. Also shown in Fig.~2 is the (exponential) distribution of the noise intensity found in the negative time ($t=-224$ ns). Apart from the uppermost curve, each of the curves is displaced by a multiple of two decades for clarity of presentation. In order to find the probability density $P(I_{\text s}/\langle I_{\text s}\rangle)$, Eq.~(12) is to be solved. However, the form of $P(I_{\text s}/\langle I_{\text s}\rangle)$ is already known \cite{Rossum,Kogan,Marin,StDy},
\begin{figure}[!]
\includegraphics[width=3.5in]{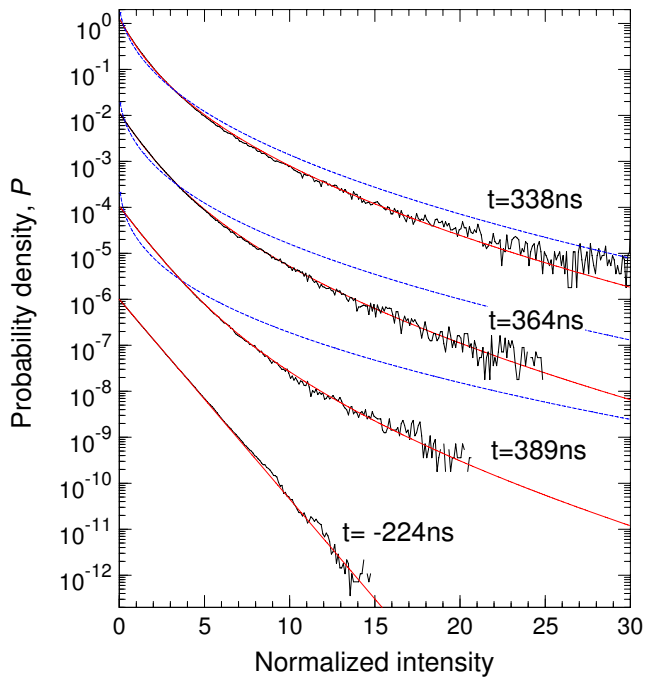}
\caption{(color on line). Probability densities of the pulsed transmitted intensities at the time delays $t=338$, 364, and 389 ns, and in the negative time ($t=-224$ ns), indicated by the vertical dashed lines in Fig.~1. The black solid curves are the measured probability densities $P(I/\langle I\rangle)$. The blue dashed curves are the probability densities $P(I_{\text s}/\langle I_{\text s}\rangle)$, deduced from Eq.~(16) using the corresponding values of $\mathrm{var}[I_{\text s}(t)/\langle I_{\text s}(t)\rangle]$ found from Fig.~1b. The red dotted curves shown through the data are $P(I/\langle I\rangle)$ deduced from Eq.~(12).}
\end{figure}
\begin{equation}
P\left(y\equiv I_{\text s}/\langle I_{\text s}\rangle\right)=\int_{0}^{\infty}\frac{dz}{z}P(z)\exp(-y/z),
\end{equation}
with
\begin{equation}
P(z)=\int_{-i\infty}^{i\infty}\frac{d\upsilon}{2\pi i}\exp{\![\upsilon z-\Phi(\upsilon)]},
\end{equation}
where
\begin{equation}
\Phi(\upsilon)=(2/3\kappa)
\ln^{2}\!\left(\sqrt{1+3\upsilon\kappa/2}+\sqrt{3\upsilon\kappa/2}\right),
\end{equation}
and $\kappa=(\mathrm{var}[I_{\text s}(t)/\langle I_{\text s}(t)\rangle]-1)/2$, that is, the probability density of the normalized transmitted intensity is given in terms of a single parameter, its variance. The values of $\mathrm{var}[I_{\text s}(t)/\langle I_{\text s}(t)\rangle]$ corresponding to the three time delays are 2.48, 2.82, and 3.42, respectively, as found from Fig.~1b. The respective probability densities $P(I_{\text s}/\langle I_{\text s}\rangle)$ are deduced from Eq.~(16) and shown as the blue dashed curves in Fig.~2. Thus we found $P(I_{\text s}/\langle I_{\text s}\rangle)$ without solving Eq.~(12). To check the validity of the $P(I_{\text s}/\langle I_{\text s}\rangle)$, we use Eq.~(12) to deduce $P(I/\langle I\rangle)$ and compare it to the measurement. The calculated $P(I/\langle I\rangle)$ are shown as the red dotted curves in Fig.~2 and are in excellent agreement with the measured densities.

\section{Conclusions}

In conclusion, we have found the transformation of the intensity statistics of random signals in the presence of additive Gaussian noise. The transformation of the intensity probability density is given by the Bessel transform of Eq.~(12). This can be solved to retrieve the intensity statistics of signal from Gaussian noise, given the average noise intensity. We used Eq.~(12) to determine the intensity statistics of pulsed microwave transmission through random media from Gaussian noise at long delay times. More generally, the results of this work can be applied to a mixture of independent random signals, one of which is a complex-valued Gaussian random process.

\begin{acknowledgments}
We thank A.Z. Genack and S. Bobkov for valuable discussions.

\end{acknowledgments}

\end{document}